\documentclass{nature}

\usepackage[T1]{fontenc}
\usepackage{textcomp}
\usepackage{amssymb}
\usepackage{amsmath}

\usepackage{gensymb}
\usepackage{caption}
\usepackage{bm}
\usepackage{lineno}
\usepackage{multibib}
\usepackage{xcolor}
\usepackage{hyperref}
\usepackage{afterpage}
\usepackage{graphics}
\usepackage{epstopdf} 

\newcites{main}{References}
\newcites{methods}{References for Methods}

\usepackage{graphicx}
\makeatletter
\let\saved@includegraphics\includegraphics
\AtBeginDocument{\let\includegraphics\saved@includegraphics}
\renewenvironment*{figure}{\@float{figure}}{\end@float}
\makeatother

\title{A jet bent by a stellar wind in the black hole X-ray binary Cygnus X-1}

\author{S.~Prabu$^{1,2}$$^{*}$, J.C.A. Miller-Jones$^{1}$$^{*}$, A. Bahramian$^{1}$,
V. Bosch-Ramon$^{3}$, S. Heinz$^{4}$, S.J. Tingay$^{1}$, C.M. Wood$^{1}$,
 A.J. Tetarenko$^{5}$, T.N. O'Doherty$^{1}$, 
 V. Tudose$^{6}$}

\begin{document}

\maketitle

\begin{affiliations}
\item International Centre for Radio Astronomy Research -- Curtin University, Perth, WA 6845, Australia
\item University of Oxford, Department of Physics, Astrophysics, Denys Wilkinson Building, Keble Road, OX1 3RH Oxford, United Kingdom
\item Departament de F\'isica Qu\`antica i Astrof\'isica,
Institut de Ci\`encies del Cosmos Universitat de Barcelona (ICCUB),
Universitat de Barcelona (IEEC-UB), E08028 Barcelona, Catalonia, Spain
\item Department of Astronomy, University of Wisconsin–Madison,
Madison, WI 53706, USA
\item Department of Physics and Astronomy, University of Lethbridge,
Lethbridge, Alberta, T1K 3M4, Canada
\item Institute of Space Science - INFLPR Subsidiary, 077125 Magurele, Romania
\end{affiliations}


\begin{abstract}
Jets provide an important channel for kinetic feedback from accreting black holes into their environment, without which models of the formation of large-scale structure in the universe fail to reproduce the observed properties of galaxies. Hence, an accurate measurement of jet power is critical for understanding black hole growth through accretion and also for quantifying the impact of kinetic feedback. However, the absence of instantaneous jet power measurements has precluded direct comparisons with the accretion luminosity, forcing kinetic feedback models to rely on ad hoc assumptions about how much jet power is released per accreted amount of mass. Here we report the detection of stellar wind-induced bending of the jets in the black hole X-ray binary Cygnus X-1, using 18 years of high-resolution radio imaging. By modeling jet-wind interactions, we determine the current kinetic instantaneous power of the jet to be log$_{10}(L_{\rm jet}/{\rm erg\,s}^{-1}) = 37.3_{-0.2}^{+0.1}$, comparable to the accretion energy determined from its bolometric X-ray luminosity. This result critically places prevailing assumptions about the energetics of black hole powered jets in both galaxy formation simulations, and in scaling models of black hole accretion, on a firm empirical footing.
\end{abstract}

Understanding the impact of accreting supermassive black holes on the evolution of galaxies and cosmic structures is one of the key motivators for studying relativistic jets (e.g., \citemain{croton:06, weinberger:17, krause:23}). Jets are observed to drive large-scale shocks, pollute interstellar gas with magnetic fields \cite{heinz2008} and cosmic rays, generate large-scale turbulence, and evacuate large-scale cavities of gas on scales of galaxy groups and clusters \cite{fabian:00,mcnamara:07}. A fundamental difficulty in modeling the kinetic feedback comes from the lack of an instantaneous jet power measurement, which would inform us of the fraction of the accreted energy that is converted into the kinetic energy of the jet. Hence, we have had to rely on time-averaged jet power estimates from calorimetry of jet-inflated bubbles to constrain this feedback, which involves averaging the total kinetic power output by a jet over its lifetime---which far exceeds the variability timescale of the accretion flow. Due to this mismatch in timescales, calorimetric measurements cannot be used to accurately calibrate the instantaneous kinetic feedback efficiency of accreting black holes, which is a key input for models of large-scale structure formation (see Methods). These reasons underscore the need to develop a method to measure the {\tt instantaneous} jet power of an accreting black hole.

High-mass black hole X-ray binaries (BHXRBs) provide a unique opportunity to measure the instantaneous jet power in an individual system via the theoretically-predicted interactions between the stellar wind and the jets \cite{yoon2015global, bosch2016effects, perucho2008}, although these have not to date been observationally confirmed. In these systems, the black hole accretes from the strong stellar wind of its massive companion star. The jets launched by the black hole must then propagate outwards through that wind. The impact of the wind can bend the jet away from the companion star \cite{perucho2008, yoon2015global}, which when combined with the orbital motion of the black hole results in a helical jet outflow \cite{bosch2016effects}. The overall jet trajectory is set by the relative strengths of the wind momentum flux and the momentum flux of the jet, hence if the wind parameters are known, instantaneous measurements of the jet power, speed, geometry, and any misalignment between the jet and the binary can be made. 

At a distance of $2.22$\,kpc \cite{miller2021cygnus}, the high-mass BHXRB Cygnus X-1 contains a $21.2 \pm 2.2$ $M_{\odot}$  black hole \cite{miller2021cygnus} in a $5.6$-day binary orbit \cite{brocksopp1999} with a supergiant companion of spectral type O and mass $40.6^{+7.7}_{-7.1}$ $M_{\odot}$ \cite{miller2021cygnus}. The black hole accretes from the stellar wind of the donor star, whose mass-loss rate is $(2.57\pm0.05) \times10^{-6} M_{\odot}$\,yr$^{-1}$ \cite{gies2003}. In the hard X-ray spectral state, a steady jet is launched from close to the black hole, which can be resolved by high angular resolution radio observations with very long baseline interferometry (VLBI) \cite{stirling2001relativistic}. Early VLBI observations of Cygnus X-1 performed in the late 1990s detected only the Doppler-boosted approaching jet, extending 10--15 milliarcseconds (mas) away from the bright core at a position angle of approximately $-26^{\circ}$ east of north \cite{stirling2001relativistic,reid2011}. A more sensitive VLBI campaign performed in 2016 detected the corresponding Doppler-deboosted receding jet for the first time \cite{miller2021cygnus}. The receding jet was found to share a similar position angle to the approaching jet on the plane of the sky. The small measured eccentricity of the binary ($e=0.019\pm0.003$ \cite{miller2021cygnus}) and low space velocity with respect to its suggested birthplace in the Cygnus OB3 association \cite{mirabel2003} imply a relatively small natal kick at black hole formation, such that the jet axis should be relatively well aligned with the binary orbit. However, recent works have suggested a substantial misalignment between the jet axis and the orbital angular momentum vector \cite{krawczynski2022polarized, zdziarski2023misalignment}. Moreover, the calorimetry of Cygnus X-1\cite{gallo2005dark,russell2007jet,sell2015} is used to anchor many observationally-derived black hole scaling relations (such as jet power vs radio luminosity\cite{heinz_grimm2005}) across the entire mass range of black holes (using the scale-invariant nature of black holes\cite{heinz_sunyaev2003,merloni:03,falcke:04}), thus further highlighting the need to measure the instantaneous jet power of Cygnus X-1. 


The 2016 VLBI campaign performed six $8.4$-GHz observations with the Very Long Baseline Array (VLBA) and three $5$-GHz observations with the European VLBI Network (EVN), monitoring the hard state jet of Cygnus X-1 around an entire 5.6-day binary orbit. We re-analysed these data, detecting both the approaching and receding jets in every observation, with a stack of the VLBA images finding a brightness ratio in the range $1.5-5.5$ (see Methods). Stacking archival VLBA observations made in both 1998 and 2009 also recovered the receding jet, at similar brightness ratios, showing that the jets are intrinsically two-sided.

\begin{figure}
\begin{center}
\includegraphics[width=\linewidth,keepaspectratio]{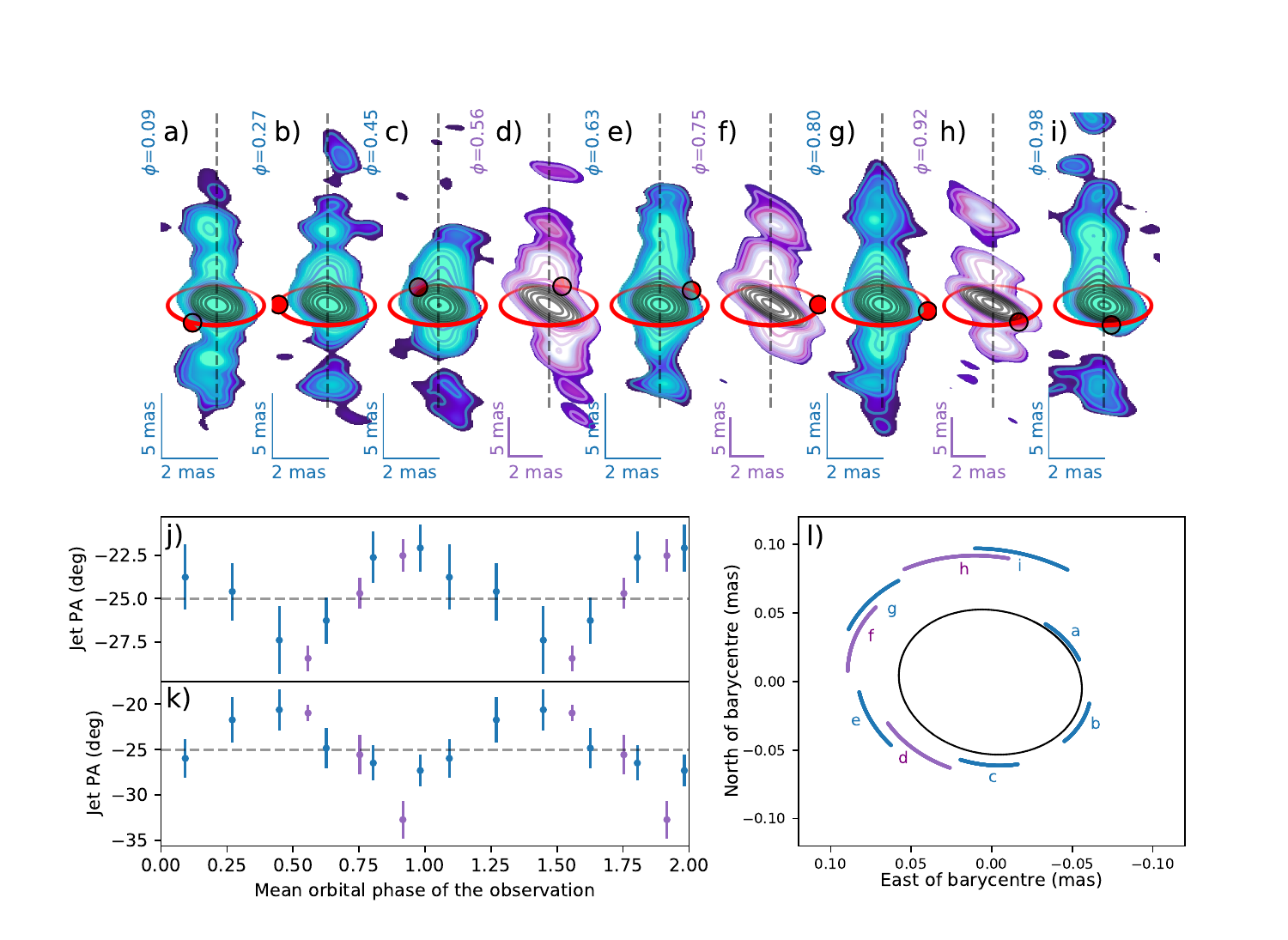}
\linespread{1.0}\selectfont{}
\caption{ \textbf{High-resolution imaging of the jets in Cygnus X-1 over a full binary orbit in 2016.}  Panels (a-i) show the individual VLBI images (with the six 8.4-GHz VLBA observations in blue and the three 5-GHz EVN observations in purple), rotated counter-clockwise by $25^{\circ}$, and with an asymmetric axis ratio to help visualise the jet bending. The donor star's orbit (scaled up by a factor of 30 for the VLBA and 50 for the EVN) is shown in each image in red. The solid red circle indicates the star's position at the mid point of the observation. The orbital phase ($\phi$) of the observation is mentioned in the corresponding  panels. (j) and (k) show the median position angle of the approaching and receding jets as measured from the core. The bending appears to lag by quarter of an orbit, due to the finite time taken by the bent jet to travel downstream. Panel l) shows the black hole orbit projected onto the plane of the sky, along with the phase coverage of each observation as shaded arcs in blue (VLBA) and purple (EVN), whose radii increase with time to avoid confusion.}
\label{FigMainPlot}
\end{center}
\end{figure}

\begin{figure}
\begin{center}
\includegraphics[width=0.8\linewidth,keepaspectratio]{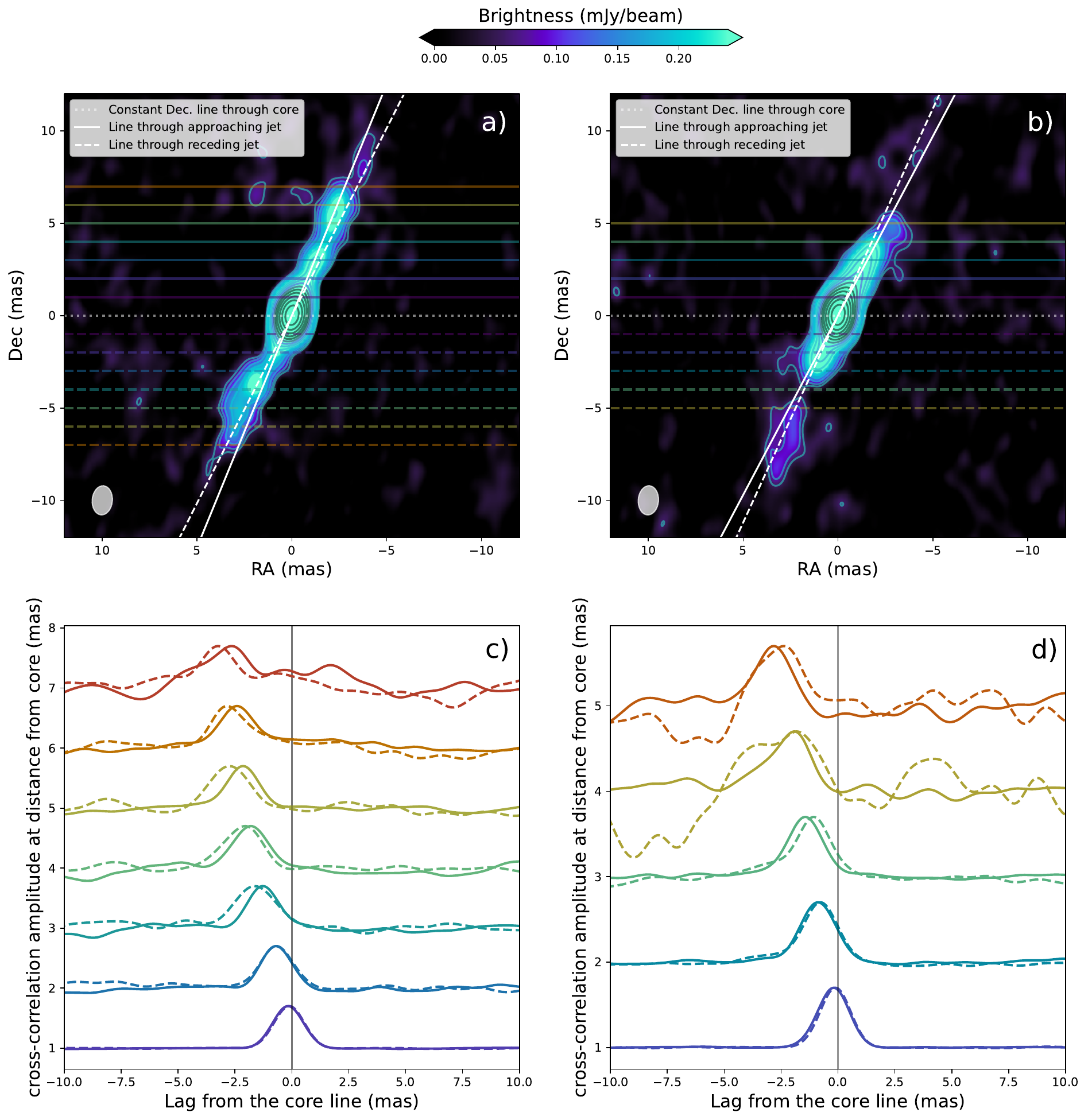}
\linespread{1.0}\selectfont{}
\caption{\textbf{A model-independent demonstration of bent jets in Cygnus X-1.} Panels (a-b) show VLBA images of Cygnus X-1 close to superior and inferior conjunction, when the jets showed significant deviations from the median position angle. The grey ellipse in each image indicates the synthesized beam size. The jet brightness profile was measured along lines of constant declination spaced by 1 mas, at the locations indicated by the horizontal lines for the core (dotted), approaching jet (solid), and receding jet (dashed).  A linear fit to the jet ridge lines is shown in white, demonstrating that the approaching (solid) and receding (dashed) jets are not co-linear. Panels (c) and (d) show normalized cross-correlations of the downstream jet brightness profiles indicated in panels (a) and (b), respectively, with the brightness profile at the declination of the core. The negative of the lags are shown for the receding jets (dashed curves), to aid comparison with their approaching counterparts (solid curves). As the jets are not oriented north-south, the peak cross-correlation lag increases on moving downstream. 
}
\label{Fig:crossCorrelation}
\end{center}
\end{figure}

Our analysis of individual VLBA and EVN images from the 2016 campaign showed the observed position angles of the approaching and receding jets to vary with orbital phase (Figure~\ref{FigMainPlot}). We verified in archival data that the variation in position angle of the approaching jet was reproducible, with orbital phase-dependent deviations from a median position angle that was constant over an 18-year period (Extended Data Figures ~\ref{FigVLBAMontage} and \ref{FigEVNMontage}). Moreover, a model-independent analysis of the jet (see Methods) showed the approaching and receding jets to bend in opposite directions (Figure~\ref{Fig:crossCorrelation}).

The observed difference in the position angles of the approaching and the receding jets cannot be explained by a simple precession of the jet axis, which would predict corresponding approaching and receding jet segments to be aligned. However, it naturally arises in a scenario in which the jets are bent away from the donor star by the impact of the stellar wind. As the wind density decreases with distance from the star, most of the bending should occur within about an orbital separation of the jet nozzle ($\sim0.1$\,mas on the plane of the sky), so that we see only the asymptotic bending angle at VLBI scales \cite{yoon2015global}. This explains the relatively linear nature of the observed approaching and receding jets, and argues that the central binary system is located close to where the radio emitting jets meet, at the bright core. This is in agreement with theoretical predictions of the photospheric distance in Cygnus X-1 \cite{zdziarski2012}. 

\begin{figure}
\begin{center}
\includegraphics[width=\linewidth]{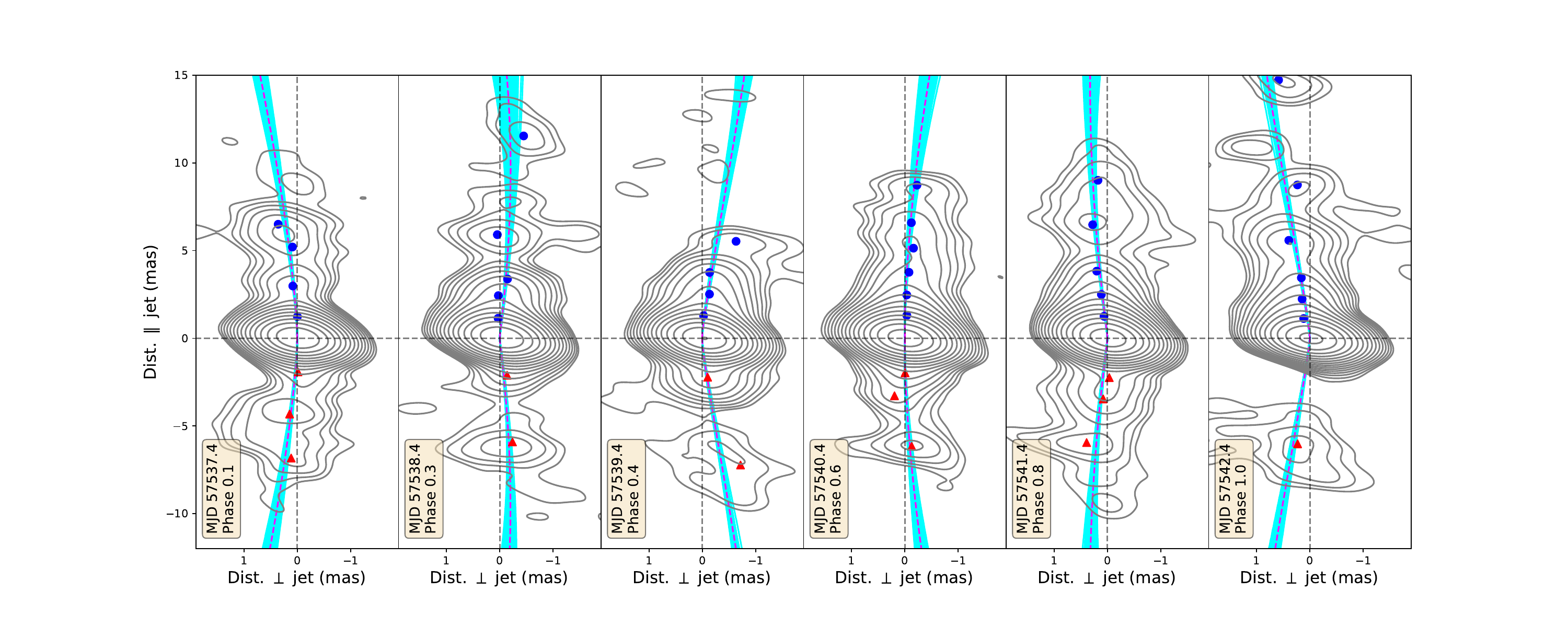}
\linespread{1.0}\selectfont{}
\caption{\textbf{Jet trajectories for each of the VLBA observations in 2016, determined from our physically-motivated jet model}. Contour plots of each image are shown, with blue and red markers denoting the locations of the point source components required to model the approaching and receding jets, respectively, determined from fitting the visibility data in the {\it uv}-plane. The errors associated with these point source components are smaller than the size of the markers. The dashed line shows the median trajectory from the results of fitting the physical model to the point source locations, and the cyan shaded area shows a set of $200$ random draws from the posterior distribution. All images have been rotated counterclockwise by 25$^{\circ}$, to align the median jet axis vertically on the figure. An asymmetric axis ratio has been used to better visualise the jet bending. The date and orbital phase of each epoch is indicated in the bottom-left corner.}
\label{Fig:TrajectoriesCase1}
\end{center}
\end{figure}

We therefore consider an analytical model of the wind-induced jet bending, balancing the wind momentum flux with the lateral momentum flux of the jet\cite{bosch2016effects, 2018A&A...618A.146M, 2022MNRAS.510.3479B}, and accounting for orbital motion to predict the overall helical structure of the jet. This physically-motivated model also accounts for non-ballistic effects as the helical jet pushes against the wind. To infer the jet properties at launch, we fit this analytical, numerically-evaluated model \cite{bosch2016effects, 2022MNRAS.510.3479B} to our measured jet structure. Simultaneously fitting all six VLBA epochs from 2016 gave a jet power of $\log_{10}(L_{\rm jet}/{\rm erg\,s}^{-1}) = 36.9_{-0.5}^{+0.2}$, a jet speed at launch of $\beta=0.47^{+0.06}_{-0.05}$, and a jet half opening angle of $0.8\pm0.5^{\circ}$. The fitted trajectories for each epoch of the 2016 VLBA observations are shown in Figure~\ref{Fig:TrajectoriesCase1}. Our fit provides the first measurement of the instantaneous jet power in any accreting black hole, along with a speed measurement for the compact jet, which has also been difficult to reliably measure (e.g., \cite{stirling2001relativistic,casella2010,tetarenko2019}).

The black hole in Cygnus X-1 was originally proposed to have formed by direct collapse \cite{mirabel2003}. In the absence of a supernova kick, the black hole spin axis would therefore be expected to be well aligned with the orbital angular momentum. However, the recent detection of high fractional X-ray polarization from the system suggested a substantial misalignment of $>18^{\circ}$ along the line of sight \cite{krawczynski2022polarized}. In contrast, a study of the frequency-dependent phase lags observed in orbital phase-folded radio light curves suggested a misalignment of $20-30^{\circ}$ in the plane of the sky \cite{zdziarski2023misalignment}. Given these constrasting claims, we explored the impact of misalignment on the calculated jet trajectories.

In the presence of a misaligned jet, we would expect highly asymmetric non-ballistic wind-jet interactions to occur near the base of the jet. Regardless of the geometry of the misalignment, it would cause the approaching jet to propagate towards the star (where the wind is denser) at a particular orbital phase, and the receding jet to do so half an orbit later. Substantial misalignment ($\gtrsim10^{\circ}$) would then lead to strong, highly asymmetric jet bending between the approaching and receding jets due to different non-ballistic forces imparted on them by the wind, which we do not observe (see Supplementary Videos). 

We re-fit our jet trajectories with additional parameters to allow for a misalignment between the jet and orbital axes, and found a jet power of $\log_{10}(L_{\rm jet}/{\rm erg\,s}^{-1}) = 37.3_{-0.2}^{+0.1}$, a somewhat higher jet speed of $\beta=0.68^{+0.08}_{-0.13}$, and a best-fitting misalignment of $5.2^{+1.0}_{-1.1}$\,degrees. However, to account for possible systematics due to averaging of the jet trajectories over a 12-hour observation (see Methods), we adopt a conservative $3 \sigma$ upper limit of $8.2^{\circ}$ on the misalignment between the jet and the binary. 

Relaxing the requirement for the jet to be conical (i.e.\ allowing the jet radius to scale with distance downstream as $r\propto z^{\epsilon}$) did not change this conclusion, giving consistent values for the jet power, jet speed, and misalignment angle, and favouring a conical jet with $\epsilon=1.02^{+0.08}_{-0.08}$. Our fitted conical jet geometry for Cygnus X-1 is in agreement with the lack of a strong re-collimation shock expected for our estimated jet power \cite{yoon2016recollimation} (see Methods).  The absence of a substantial jet-orbit misalignment is in agreement with a lack of any observed Lense-Thirring precession of the jet, as suggested by the stable mean position angle of the jet on the plane of the sky over the 18 years of VLBI observations used here (Extended Figure ~\ref{FigJetPA}). 

Such a small jet-orbit misalignment implies that alternative explanations are required for the high fractional X-ray polarization observed by IXPE, such as the presence of a relativistic outflow in the corona \cite{poutanen2023polarized}. The radio phase lags found by \cite{zdziarski2023misalignment} could be explained by the helical jet structure created by the bent jet and the orbital motion of the black hole. The low misalignment is also in agreement with the low eccentricity and peculiar velocity of the system \cite{miller2021cygnus,mirabel2003}, and with theoretical expectations for the formation of such massive black holes \cite{fryer2012}. 

Although our jet power estimate from the three physical models (without misalignment, with misalignment, and with misalignment and non-conical jet) are in agreement with each other at the $1\sigma$ level, and relatively insensitive to uncertainties in the mass-loss rate of the donor star wind (see Methods). We favour the jet power estimate from the model that allows for both misalignment and a non-conical jet geometry , i.e, $\log_{10}(L_{\rm jet}/{\rm erg\,s}^{-1}) = 37.3_{-0.2}^{+0.1}$, as it makes the smallest number of assumptions. This implies that over the few-Myr lifetime of the BHXRB, the total kinetic feedback from the jets in Cygnus X-1 would be several times $\sim10^{50}$\,erg, comparable to the kinetic feedback from a supernova. Our measured instantaneous jet power is in excellent agreement with the time-averaged jet power of $4 - 14 \times 10^{36}$\,ergs\,s$^{-1}$ derived for Cygnus X-1 through calorimetry (see Methods for extended discussion on calorimetry), and is shown in Figure ~\ref{FigJetPower}. The striking similarity between the accretion power (as infered from the X-ray bolometric luminosity of Cygnus X-1 in the hard state \cite{wilms:06}, rescaled to an updated distance of $2.22$\,kpc\cite{miller2021cygnus}) and our jet power measurement, validates the commonly assumed accretion to jet energy conversion fractions assumed in typical galaxy formation simulations\cite{croton:06,weinberger:17,pillepich:18,dave:19,byrne:24,ni:24,huvsko:24} (see Methods).

With a robust, instantaneous measurement of the jet power in Cygnus X-1, we validate the use of calorimetry for calibrating the fraction of accretion energy converted into kinetic energy of the jet. The consistency between the instantaneous and time-averaged jet power implies long term stability of the jets from radiative inefficient flows (i.e, in the hard X-ray spectral state), and vindicates the use of calorimetry to normalize the scaling relations of black holes. On larger scales, our measurement of instantaneous accretion to jet energy conversion fraction lends strong support for the assumed energy budget of accreting black hole in large-scale cosmological simulations.

\bibliographystylemain{naturemag}
\bibliographymain{sample}


\begin{figure}
\begin{center}
\includegraphics[width=\linewidth,keepaspectratio]{figures/jetpower.jpg}
\linespread{1.0}\selectfont{}
\caption{\textbf{The instantaneous jet power of Cygnus X-1 during our 2016 observing campaign, as compared to estimates of the time-averaged jet power.} We show probability density functions (PDFs) estimated from posterior samples for the instantaneous jet power for our three different models; Model 1, with no jet-orbit misalignment and conical jet geometry (blue); Model 2, allowing for misalignment with a conical jet geometry (magenta); and Model 3, allowing for both misalignment and a non-conical geometry (orange). The PDF of the derived jet power is relatively similar across all three models, and shows that the instantaneous jet power is consistent with the time-averaged power required to inflate the arcminute-scale radio nebula surrounding the source (shown by the grey shaded regions \cite{gallo2005dark, russell2007jet, sell2015}.}

\label{FigJetPower}
\end{center}
\end{figure}

\clearpage

\begin{addendum}
 \item[Supplementary Information] is linked to the online version of the paper at www.nature.com/nature.

 \item We thank Dave Russell, Adam Ingram and Chris Done for discussions. The National Radio Astronomy Observatory is a facility of the National Science Foundation operated under cooperative agreement by Associated Universities, Inc. This work made use of the Swinburne University of Technology software correlator, developed as part of the Australian Major National Research Facilities Programme and operated under licence. The European VLBI Network is a joint facility of independent European, African, Asian, and North American radio astronomy institutes. Scientific results from data presented in this publication are derived from the following EVN project code(s): RT013. This work was supported by resources provided by the Pawsey Supercomputing Research Centre with funding from the Australian Government and the Government of Western Australia. SP acknowledges Breakthrough Listen (funded by the Breakthrough Prize foundation) for their Breakthrough Listen fellowship. VB-R acknowledges financial support from the Spanish Ministry of Science and Innovation under grant PID2022-136828NB-C41/AEI/10.13039/501100011033/ERDF/EU and the María de Maeztu award to the ICCUB CEX2024-001451-M, and from the Generalitat de Catalunya through grant 2021SGR00679. V.B-R. is Correspondent Researcher of CONICET, Argentina, at the IAR. CMW and TNOD acknowledge financial support from the Forrest Research Foundation Scholarship and the Australian Government Research Training Program Scholarship. CMW also acknowledges financial support from the Jean-Pierre Macquart Scholarship. AJT acknowledges that this research was undertaken thanks to funding from the Canada Research Chairs Program and the support of the Natural Sciences and Engineering Research Council of Canada (NSERC; funding reference number RGPIN-2024-04458). VT acknowledges support
from the Romanian Ministry of Research, Innovation and Digitalization
through the Romanian National Core Program LAPLAS VII – contract no.
30N/2023. 

 \item[Author contributions] SP analyzed all archival VLBA and EVN observations of Cygnus X-1. We independently confirmed the observed bending of the jets in the 2016 VLBA campaign using JCAM-J's previously-published analysis of the BM429 observations \cite{miller2021cygnus}. JCAM-J, SP, and AB led the manuscript development. AB developed the lag correlation analysis that demonstrated the jet bending using a model-independent technique. SP and AB developed all the MCMC models used in this work. VBR led the development of the physical model to describe the jet bending, which was further modified by SP and JCAM-J to incorporate spin-orbit misalignment.  SH led the discussions on implications of the instantaneous jet power measurement and the scaling relationship between radio luminosity and jet power. SP, JCAM-J, SJT, and CMW led the model fitting analysis of the VLBI data using difmap and UVMULTIFIT. AJT performed the radio timing ana  lysis described in Methods. TNOD performed the VLA data processing. VT designed the EVN observation campaign. All authors provided input and comments on the manuscript.
 
 \item[Reprints] Reprints and permissions information is available at www.nature.com/reprints.
 
 \item[Competing Interests] The authors declare that they have no competing financial interests.
 
 \item[Correspondence] Correspondence and requests for materials should be addressed to\\
 S.P~(email: steve.prabu@physics.ox.ac.uk) and J.C.A.M.-J.~(email: james.miller-jones@curtin.edu.au).
\end{addendum}

\newpage

\begin{methods}
\subsection{Data reduction.}
The primary observations considered in this work were obtained using the VLBA and the EVN, and covered a single binary orbit of Cygnus X-1 in May--June 2016 \cite{miller2021cygnus}. To probe the long-term evolution of the jets, we also analysed all existing archival data obtained at frequencies between 5 and 8.4\,GHz \cite{stirling2001relativistic,reid2011,rushton2012weak} (in the hard state), which are tabulated in Extended Data Table ~\ref{tab:obs}. 

All VLBA observations were processed using the Astronomical Image Processing System ({\tt AIPS}), version 31DEC22 \citemethods{Greisen2003}, following the standard procedures for phase-referenced experiments\footnote{\url{http://www.AIPS.nrao.edu/cook.html}}. Having performed the recommended initial calibration steps, we then exported the Cygnus X-1 data to {\tt difmap} \citemethods{1997ASPC..125...77S} for further calibration (phase only followed by joint amplitude and phase self-calibration) and imaging. For EVN observations, we followed the EVN guide\footnote{\url{https://www.evlbi.org/evn-data-reduction-guide}}. We used the flagging and the delay, bandpass, and a-priori amplitude calibration solutions generated by the EVN pipeline, followed by global fringe fitting performed in {\tt AIPS}. The Cygnus X-1 data was then exported to {\tt difmap} for  self-calibration and imaging as done for the VLBA data. A montage of all naturally weighted VLBA (EVN) images is shown in Extended Data Figure~\ref{FigVLBAMontage} (\ref{FigEVNMontage}).

We then used {\tt difmap} to model the jet as a series of individual components. As we expect the core to be compact but resolved parallel to the jet (due to possible ejection of new components), we modeled it using an elliptical Gaussian. The downstream emission was unresolved in the transverse direction, and hence we modeled it as a series of delta functions, the number of which was determined by the extent of the resolved jet. Having generated a preliminary model in {\tt difmap}, we conducted our final model fitting using the {\tt UVMULTIFIT} package \citemethods{uvmultifit} within  the Common Astronomy Software Application ({\tt CASA} \citemethods{casa}), to determine the uncertainties associated with each of the fitted parameters.

\subsection{Model-independent illustration of a non-co-linear jet in Cygnus X-1.}
We selected two of our 2016 VLBA epochs taken at different points in the orbit, that showed significant jet bending (A and C of Figure~\ref{FigMainPlot}, respectively). We extracted intensity profiles at constant declination, along the core and every $\pm1$\,mas downstream (see top panels of Figure~\ref{Fig:crossCorrelation}). The downstream intensity profiles were cross-correlated with the intensity profile of the core, such that the lag represents the shift in right ascension of the peak intensity at the relevant declination, relative to the right ascension of the core. We show the lag measurements in the bottom panels of Figure~\ref{Fig:crossCorrelation}, inverting the lags measured along the receding jet for ease of comparison. For a co-linear jet (i.e., no bending), the lags measured at equal angular separations from the core along the approaching and receding jets would be expected to be equal in magnitude, contrary to what is shown in Figure~\ref{Fig:crossCorrelation}. For a jet aligned purely north-south, the cross-correlation in every declination slice would peak at zero lag. For any other orientation, the cross-correlation lag would increase linearly on moving downstream. On one side of the orbit (right panels), the lags along the approaching jet are larger in amplitude compared to those measured at the same distance downstream along the receding jet. This implies that the approaching jet is rotated further clockwise on the plane of the sky than the receding jet, confirming that the jets are not co-linear. On the other side of the orbit (left panels), the opposite lag behavior is seen, demonstrating that this bending of the jets varies with orbital phase.

\subsection{Physical model of the jet bending.}
Having confirmed the existence of jet bending via the model-independent cross-correlation method, we model the observed bending by considering the momentum transferred from the wind to the jet \cite{bosch2016effects, 2018A&A...618A.146M, 2022MNRAS.510.3479B}. In a non-rotating system, the wind would bend the jet away from the donor star, eventually reaching an asymptotic bending angle \cite{yoon2015global}. However, in a rotating binary system, the combination of orbital motion and radial bending would result in a helical jet outflow \cite{bosch2016effects, 2018A&A...618A.146M, 2022MNRAS.510.3479B}. 

We use a framework adapted from \cite{2018A&A...618A.146M} to calculate the jet bending, using a co-rotating frame as shown in Extended Data Figure~\ref{Fig:diagram}. The $x$-axis is pointed towards the black hole, the $z$-axis is parallel to the binary's orbital angular momentum vector, and the $y$-axis is orthogonal to both, forming a Cartesian co-ordinate system. We then define the jet's initial momentum along the $x$, $y$, and $z$ directions as
\begin{equation}
    \begin{array}{l}
        p_{{\rm jet,} x} = 0, \\
        p_{{\rm jet,} y} = 0, \\
        p_{{\rm jet,} z} = p_{\rm jet} = \frac{L_{\rm jet} \Gamma \beta}{(\Gamma -1) c},\\
    \end{array}
\end{equation}
where $L_{\rm jet}$ is the jet power, $\beta$ the jet speed at launch, $\Gamma$ is the corresponding Lorentz factor, and $c$ is the speed of light. The trajectory of the jet downstream is numerically calculated using the equation $\vec{p}_{i+1} = \vec{p}_{i} + \vec{F}_{i}dt$, where the thrust imparted by the wind ($\vec{F}_{i}$) on the jet that is transferred to the $(i+1)$th jet segment, is calculated using the wind density and wind velocity calculated upstream. Very close to the base of the jet, non-ballistic jet-wind interactions determine the jet structure, and further downstream, due to reduced wind density, a ballistic propagation of the jet is assumed by our model. An example jet trajectory calculated at different orbital phases is shown in Extended Data Figure~\ref{FigJetWindMomentum}. 

We used dynamical nested sampling \citemethods{skilling2006} implemented using the {\tt dynesty} Python package \citemethods{2020MNRAS.493.3132S} to fit our model to the individual delta components from modelling the 2016 VLBA observations of Cygnus X-1. As an initial fit, denoted as Model 1, we used a minimal subset of free parameters, comprising the jet speed at launch $\beta$, the jet power $L_{\rm jet}$, the jet half-opening angle near the base $\phi_{1/2}$, the inclination angle of the jet at launch $i$, the donor star wind velocity $v_{\rm wind}$, and the jet $PA$ at launch projected on the plane of the sky. The terminal velocity of the donor star wind is expected to be within the range $1600-2400$\,km\,s$^{-1}$, \cite{gies2008, grinberg2015} and hence is left as a free parameter in our fit with appropriate bounds. The priors and posterior distributions of the model fit are tabulated in Extended Data Table ~\ref{tab:modelfits}. We found a jet power of $\log_{10}(L_{\rm jet}/{\rm erg\,s}^{-1})=36.94_{-0.49}^{+0.22}$, a jet launch speed of $\beta=0.47_{-0.05}^{+0.06}$, a half-opening angle of the jet of $0.8\pm0.5^{\circ}$, a jet launch position angle of $-25.0\pm0.1^{\circ}$, and a wind velocity of $\le 1870$\,km\,s$^{-1}$.

Recent studies have raised the possibility of a misalignment between the jet and the binary orbit in Cygnus X-1, whether along the line of sight \cite{krawczynski2022polarized}, or in the plane of the sky \cite{zdziarski2023misalignment, 2024ApJ...962..101Z}. A misaligned jet would result in orbital phase dependent asymmetric bending of the approaching and receding jet, as one of the jets would be launched towards the star, undergoing significantly greater bending due to strong non-ballistic interaction with the denser wind. Hence, we modify our physical model to evaluate the possibility of a misalignment, by using two additional parameters to describe the initial orientation of the jet at launch, namely the misalignment angle $\theta$, and the binary orbital phase, $az$, when the misaligned jet points along the line joining the star and the black hole. For an $az$ value of 0 (or 0.5), the jets would be pointed towards/away from the star at both superior and inferior conjunction (i.e, a misalignment along the line of sight direction), and for a value of $az=0.25$ (or $0.75$), the jets would be pointed towards/away from the star at quadrature phases (a misalignment in the plane of the sky). Also, as the jet is evaluated in a co-rotating frame, in the presence of a misalignment, each downstream segment would have been launched at a different orbital phase, resulting in different initial jet momentum along the x and y directions (in the case of no-misalignment, the initial momentum of all downstream jet segments are zero along the x and the y directions). Hence, in our numerical model, each downstream jet segment is mapped to the orbital phase at which it was launched, which in turn determines the initial momentum of the jet segment along the $(x,y,z)$ directions. For an assumed jet segment length $dl$, an orbital period $T_{0}$, and a jet speed $\beta$, then each downstream jet segment (indexed as $n=0,1,2,3,...$, where $n$ is the segment number with $n=0$ being the segment right next to the black hole) would have been launched at an earlier orbital phase $n\times \Delta \phi$, where $\Delta \phi = dl/(\beta T_{0})$. This offset in phase is then used to determine the jet segment's initial momentum at launch. If the black hole is currently at an orbital phase $\phi_{0}$, then the initial jet momentum for the $n$th downstream segment in the co-rotating frame is given by,
\begin{equation}
    \begin{split}
        p_{{\rm jet},n,x} &= p_{\rm jet} \cos(\phi) \sin(\theta), \\
        p_{{\rm jet},n,y} &= p_{\rm jet} \sin(\phi) \sin(\theta), \\
        p_{{\rm jet},n,z} &= p_{\rm jet} \cos(\theta),
    \end{split}
\end{equation}
where $\phi=2\pi(\phi_{0} -n \Delta \phi + az)$, and $\phi_{0}$ is the binary phase of the epoch being considered. Animations of jet trajectories for a misaligned jet are provided in the Supplementary section. Fitting this misaligned jet model (denoted as Model 2) to the data points from the 2016  VLBA observations found a comparable jet power of $\log_{10}(L_{\rm jet}/{\rm erg\,s}^{-1})=37.28_{-0.23}^{+0.13}$, a jet speed of $\beta=0.68_{-0.11}^{+0.07}$ at launch, a half-opening angle of $1.8^{\circ}\pm0.7^{\circ}$, a misalignment of $5.2_{-1.1}^{+1.0}$ degrees with a phase of $az=0.34_{-0.02}^{+0.03}$ (approximately halfway between an orientation along the line of sight and in the plane of the sky), and a wind velocity of $\le 1980$\,km\,s$^{-1}$. The adopted priors and resulting posteriors are tabulated as Model 2 in Extended Data Table \ref{tab:modelfits}. We also note that Model 2 finds a mean jet position angle of $-27.8_{-0.4}^{+0.5}$ degrees. We expect this to be the position angle of the jet at launch, and when combined with the asymmetric jet bending from the $5.2^{\circ}$ jet-orbit misalignment, results in the apparent large-scale VLBA jet appearing to precess about a mean position angle of $-25^{\circ}$, in agreement with the value from the scenario with no misalignment considered in Model 1. 

Our final model (Model 3) incorporated non-conical jet geometries into Model 2. We allowed the jet cross-section to scale with distance downstream as $r\propto z^{\epsilon}$, and fit for the parameter $\epsilon$ (where $\epsilon=1$ for a conical jet). The fit found the jet power to be consistent with values from the previous models, and $\epsilon=1.0\pm0.1$, which is consistent with a conical geometry within uncertainties. The fit parameters are tabulated as Model 3 in Extended Data Table~\ref{tab:modelfits}. The median and 200 random draws from the posterior distributions from Model 1/2/3 are shown overlaid on top of the six VLBA images from 2016 in the top/middle/bottom row of Extended Data Figure~\ref{Fig:allFitsTrace}. 

\subsection{Misalignment.} The presence of a small but non-zero misalignment of $5.2_{-1.1}^{+1.0}$ degrees is consistent with the maximum plausible misalignment of $\sim10^\circ$ inferred from the ratio of the system's kick velocity and its pre-supernova orbital velocity \cite{miller2021cygnus}. The small but non-zero binary eccentricity and space velocity would also be consistent with a small but non-zero misalignment. However, we note the 12-hour duration of the individual VLBA epochs, which would have averaged the jet trace over a non-negligible fraction of the 5.6-day binary orbit, could have affected the observed trajectories of our VLBI images. We therefore retain a degree of caution in interpreting the misalignment fit, and prefer to draw a more conservative conclusion, ruling out misalignments $>8^{\circ}$. This is far smaller than the 20--30$^{\circ}$ misalignments inferred by \cite{krawczynski2022polarized,zdziarski2023misalignment}, and in agreement with the stable mean position angle of the jet observed during the 18 years of archival data (see Extended Data Figure~\ref{FigJetPA}).

\subsection{Re-collimation shocks.} For sufficiently low jet power (below a critical value) in HMXBs, a lateral re-collimation shock is expected to be driven into the jet by the wind\cite{yoon2016recollimation}, resulting in non-conical jet geometry. For a given donor star mass loss rate ($\dot{M_{\rm w}}$), wind velocity ($v_{\rm w}$), and jet velocity ($v_{\rm j}$), the critical power is given by $P_{\rm cr} = \dot{M_{\rm w}}  v_{\rm w}  v_{\rm j}/16$ \cite{yoon2016recollimation}. Using the binary parameters adopted in this paper, we find the critical jet power to be $(2.5-5.9)\times10^{37}$ erg\,s$^{-1}$, comparable to our estimate of instantaneous jet power. Thus our determined conical jet (i.e, lack of a strong re-collimation shock) from Model 3 is in perfect agreement with theoretical expectations. 

\subsection{Brightness and spectral index of the jet.}
The brightness ratio of the approaching and receding jets, combined with the spectral index, constrains the intrinsic jet speed\citemethods{RyleAndLongair1967, Scheuer1979}. From the VLBA and EVN montage (Extended Data Figures \ref{FigVLBAMontage} and \ref{FigEVNMontage}), we note that the receding jet is only detected in epochs with noise levels below $0.3$ mJy beam$^{-1}$. To accurately measure the brightness ratio and confirm that the non-detection of the receding jet is solely due to noise, we create stacked images using temporally close observations with similar setups (e.g., bandwidth). After correcting for parallax, orbital, and proper motion shifts, we produce three stacks: one from 1998 (BS060AX, BS060BX, BS060CX), one from 2009 (BR141A, BR141B, BR141C, BR141D), and one from 2016 (BM429B–BM429G), shown in Extended Data Figure~\ref{Fig:brightnessRatio} (left panels). Stacking revealed the receding jet in epochs where it was previously undetected.

With the receding jet confirmed as a persistent feature over $\sim 2$ decades, we measure the brightness ratio in the three stacked images (1998, 2009, 2016). We extract spatially resolved intensity profiles every $0.5$ mas along both jets—approximately half the restoring beam size—and fit 1D Gaussians to these profiles. The measured intensities and brightness ratio are presented in the right and bottom panels of Extended Data Figure~\ref{Fig:brightnessRatio}. We infer a conservative brightness ratio range of 1.5–5.5, which we use to constrain the jet speed in the next section. 

During the 2016 VLBI observations that monitored a full binary orbit (May–June 2016), the Karl G. Jansky Very Large Array (VLA) simultaneously observed Cygnus X-1 under program VLA/15B-236. We reduced the data following NRAO's VLA guide\footnote{\url{https://casaguides.nrao.edu/index.php/VLA_Continuum_Tutorial_3C391-CASA5.4.0}}. The spectral index (defined as $S_{\nu}\propto \nu^{\alpha}$, where $S_{\nu}$ is the flux density measured at a given frequency $\nu$), derived from flux densities at 5.25 and 7.45 GHz for each epoch, is listed in Table ~\ref{tab:vla}. Since the spectral index varies with orbital phase due to changes in path length through the stellar wind, we use the measurement closest to inferior conjunction (2016 June 1) as indicative of the jet's intrinsic spectral index.

\subsection{Hard state jet speed of Cygnus X-1.}
\label{sec:hardstatejetspeed}
The brightness ratio of intrinsically symmetric jets at equal angular separations from the core constrains the product $\beta\cos i$, where $\beta$ is the jet speed and $i$ is the system’s inclination \citemethods{RyleAndLongair1967, Scheuer1979}. For a continuous jet as seen in Cygnus X-1, the jet speed depends on the measured brightness ratio ($S_{\rm ratio}$) as:
\begin{equation} 
\beta = \frac{1}{\cos i} \left[\frac{S_{\rm ratio}^{1/(2 - \alpha)} - 1}{S_{\rm ratio}^{1/(2 - \alpha)} + 1}\right], \label{eq1}
\end{equation}
where $\alpha$ is the spectral index ($S_{\nu}\propto \nu^{\alpha}$). Given the binary inclination of $27.5^\circ \pm 5^\circ$ \cite{miller2021cygnus}, accounting for a possible misalignment $<8^\circ$, we use a brightness ratio of 1.5–5.5 and a spectral index of $0.055\pm0.016$ to constrain the hard state jet speed from the brightness ratio to be $\beta = 0.27 - 0.46$ ($68\%$ high-density interval).

Irrespective of the model used, we find the hard state jet speed of Cygnus X-1 to be only mildly relativistic, with $\beta <0.76$ within the 68\% credible intervals. The physically-motivated model finds a jet speed of $\beta=0.47^{+0.06}_{-0.05}$ for a jet aligned with the binary (Model 1), and $\beta=0.68_{-0.13}^{+0.08}$ in the presence of a misalignment (Model 2). A re-run of the radio timing study presented in \citemethods{alexRadioTiming}, incorporating updated constraints on the photosphere distance\cite{{zdziarski2023misalignment}}, finds $\beta=0.42 - 0.64$ ($68\%$ high density interval), providing further confidence in the speeds obtained from our physical model.

The jet speed determined from the brightness ratios is in marginally better agreement with the fitting results from physical Model 1 than from Models 2 and 3. This could be taken to favor an aligned jet in Cygnus X-1, but uncertainties associated with the brightness ratio measurement (such as 
image based convolution effects in the VLBI images, or physical effects such as mass loading, or a radio emission contribution from a slower jet sheath where much of the wind-jet interaction occurs) limits its discriminative power.  However, such topics remain beyond the scope of the current paper, and impossible to probe observationally with the current data.

Regardless of the presence or absence of a misalignment in Cygnus X-1, based on our physical modeling we find the jet speed to lie in the range $\beta=0.42-0.76$. A more precise determination of the jet speed in Cygnus X-1 would require higher-precision measurements of the jet trajectory, on both smaller and larger angular scales.

{\subsection{Measurements and applications of jet power.}
\label{sec:power}
Measurements of jet power are key input for theories of accretion as well as models of galaxy formation and evolution, as both require knowledge of the partitioning of accretion energy between advection, radiation, and jet/kinetic power. However, due to the featureless non-thermal synchrotron spectra, direct measurements of jet power from observations of jets are difficult. Neither the jet velocity nor composition can be determined directly from observations of the synchrotron emission alone, and jet power estimates therefore rely on ad-hoc assumptions about composition and the equipartition fraction of the plasma, as well as the bulk Lorentz factor. These assumptions introduce uncertainties of at least an order of magnitude.

Observational constraints on the time-averaged jet power can be derived from modeling the interactions between the jets and the surrounding ISM \cite{tetarenko2018power, tetarenko2020power}

The gold standard for measuring jet power to date relies on calorimetric methods, whereby the interaction of jets with some known target material is used to determine the average jet power (e.g, \cite{tetarenko2018power, tetarenko2020power}). The most common method uses observations of jet-inflated cavities in galaxies, groups, and clusters of galaxies\cite{mcnamara:07} (this method was also used when deriving the jet power required to inflate the nebula observed around Cygnus X-1\cite{gallo2005dark}). These estimates have been used to derive scaling relations between the radiative and the kinetic power of large samples of jets \citemethods{birzan:08},\cite{cavagnolo:10}.

Because these relations rely on measurements of the kinetic power {\em averaged} over the dynamical timescale of the cavity, they introduce large uncertainties when used to estimate {\em instantaneous} jet powers, given the generally rapid variability in accretion and therefore jet power inferred by observations of X-ray binaries, blazars, and nearby jets.

Because of these large uncertainties, numerical models of structure formation generally leave the jet power as a free parameter $\eta_{\rm jet} = P_{\rm jet}/\dot{M}c^2 \sim 0.1$ \cite{weinberger:17,pillepich:18,dave:19,byrne:24,ni:24,huvsko:24} that is then tuned so that the simulations reproduce the observed properties of galaxies. While this assumption appears reasonable (given its order-of-magnitude agreement with the maximum radiative luminosity achievable for radiatively efficient accretion\citemethods{novikov:73}), it could be substantially strengthened by direct observational verification of the value of $\eta_{\rm jet}$, which underscores the importance of the instantaneous jet power measurement of Cygnus X-1 presented here, which is broadly consistent with the findings of the aggregate jet power needed in cosmological simulations.}


\subsection{Implications for high energy emission from Cygnus X-1.}
The LHAASO detection of Cyg X-1 up to ~60 TeV\cite{lhaaso2024} implies a
luminosity in gamma rays of $\sim 1\times 10^{32}$ erg\,s$^{-1}$ in the range 10--60 TeV. As proposed by the LHAASO study, we investigate the possibility of the observed high energy emission to originate from the jet interaction with the shell surrounding Cygnus X-1. Firstly, under the premise that the observed TeV emission is from inverse comptonisation (IC) of ambient microwave and far infrared photons by the electrons accelerated by the jet termination within the bubble, the IC cooling time-frame ($t_{IC} \sim 2\times10^4 \cdot u^{-1}_{\text{FIR}} \cdot E^{-1}_{100\text{TeV}}$, assuming a target photon energy density of  $\sim 3\times 10^{-13}$ erg\,cm$^{-3}$) is comparable to the $\sim 20$\,kyr age of the bubble\cite{gallo2005dark}. Secondly, as our obtained jet power is large enough to power the observed gamma-ray luminosity (sufficiently conservative to account for some inefficiency in the conversion of jet energy into non-thermal electrons, and from the latter into gamma rays $\>10$\, TeV), our measured jet power favours the proposed hypothesis by LHAASO that the observed TeV emmision in Cygnus X-1 originates from the jet interaction with the bubble surrounding it. 

\subsection{Calorimetry of Cygnus X-1.}
Of the four calorimetric studies of Cygnus X-1 in the literature \cite{gallo2005dark, russell2007jet, sell2015, atri:25}, the most recent study using MeerKAT \cite{atri:25} differs in their average jet power measurement by over an order of magnitude. We attribute this discrepancy to the difference in shock velocities considered in the different studies. The MeerKAT study considers a wide range of shock velocities between $21-364$, km/s using constraints from the observed radio emission and upper limits from lack of observed X-ray emission. \cite{russell2007jet} further rules out shock velocities $<100$\,km/s using the ratios of the observed O[{\sc iii}] and H$\alpha$ emission lines. Using constraints from radio, upper limits from X-ray, and a series of optical emission lines, \cite{sell2015} places a very tight constraint on the shock velocity to be $220^{+50}_{-30}$ km/s. As the jet power in calorimetric models scale as the third power of the shock velocity, we attribute the large discrepancy between \cite{atri:25} and the remaining studies to be due to their consideration of very small shock velocities. Due to this reason, average jet power estimates from \cite{atri:25} is not discussed in the main article. 

\subsection{Donor star mass-loss rate.}
Recent results\cite{Lai24, Ramachandran25} have suggested a higher uncertainty in the donor star mass-loss rate than inferred from the originally-accepted value\cite{gies2003}. We therefore tested the sensitivity of our results to the assumed wind mass-loss rate of the donor star, re-running our Model 1 fit for mass-loss rates in the range $0.5$--$7\times10^{-6} M_{\odot}$\,yr$^{-1}$ \cite{Lai24, Ramachandran25}. As shown in Figure~\ref{masslossrate}, we found that only the jet half-opening angle varies significantly to match the momentum rate of the wind, with the fitted opening angle increasing at lower mass-loss rates and decreasing at higher rates. This demonstrates that our inferred jet power is relatively insensitive to the assumed wind mass-loss rate from the donor star.

\begin{figure}[h!]
\centering
\renewcommand{\figurename}{\textbf{Extended Data Figure}}
\setcounter{figure}{0}  
\includegraphics[width=0.85\textwidth]{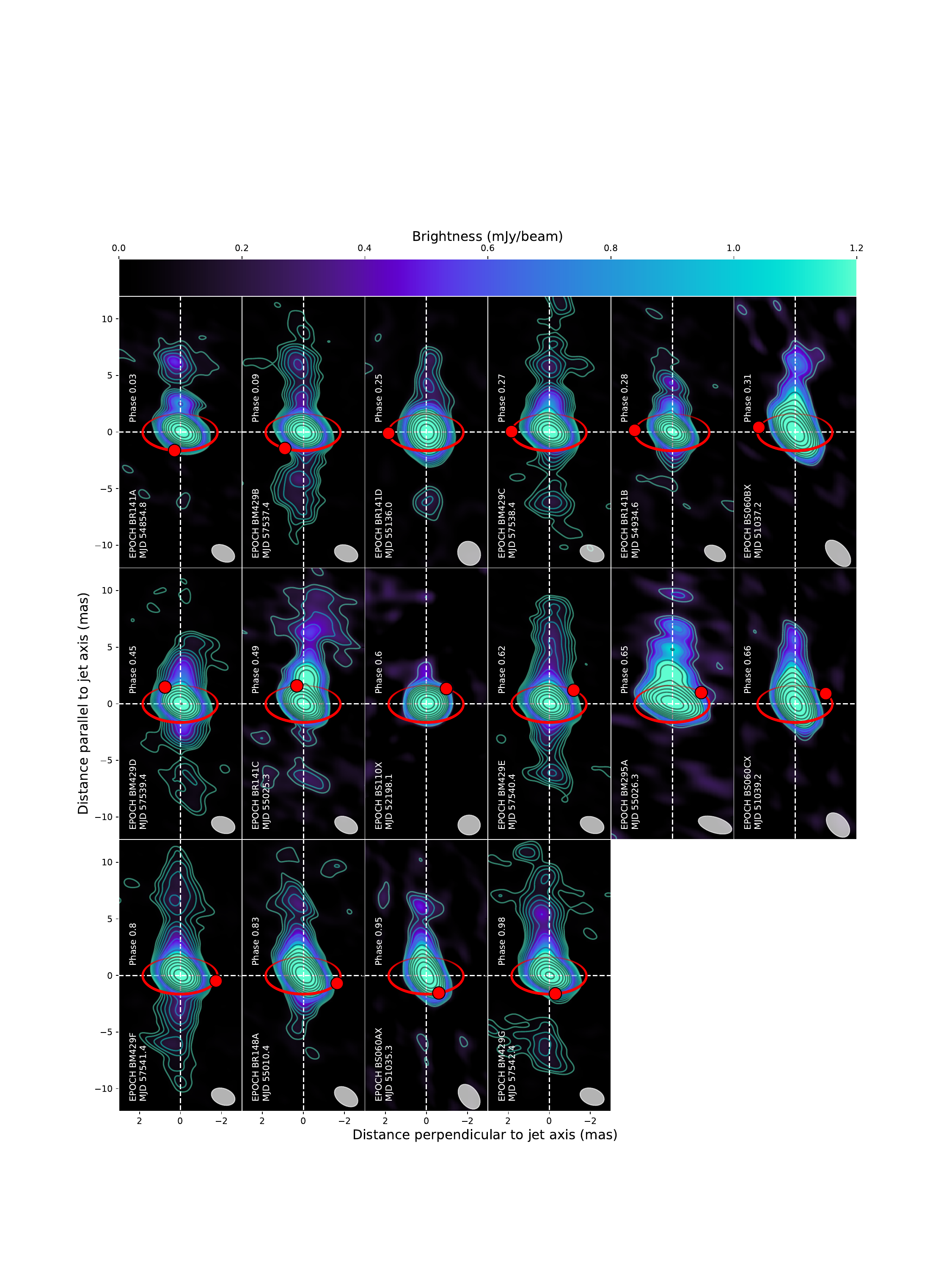}
\linespread{1.0}\selectfont{}
\caption{\textbf{A montage of all 8.4-GHz VLBA observations, sorted by orbital phase.} The contours are drawn at $\sigma \times \sqrt{2}^{n}$, where $n=3, 4, 5, ...$ and $\sigma$ is the image noise, as tabulated in Table~\ref{tab:fluxdensities}. The red ellipse shows the orbit of the donor star as seen by the black hole, scaled up by a factor of 30 for visualisation purposes.  The solid red circle shows the star's location at the mid-point of the observation. The grey ellipse in the bottom right of each panel shows the size of the restoring beam. The images have been rotated by $25^{\circ}$ counter clockwise.}\label{FigVLBAMontage}
\end{figure}

\begin{figure}[h!]
\centering
\includegraphics[width=1.0\textwidth]{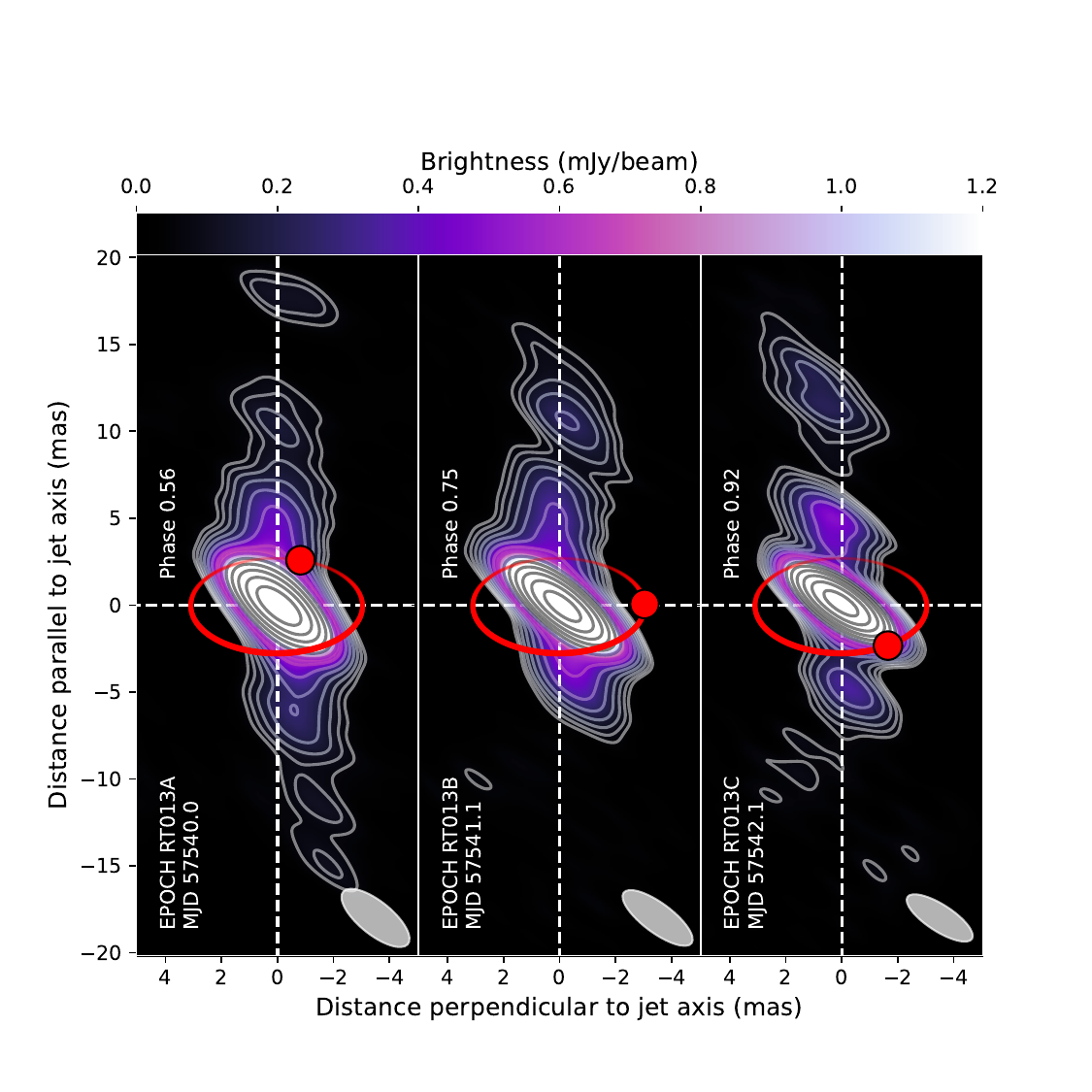}
\renewcommand{\figurename}{\textbf{Extended Data Figure}}
\linespread{1.0}\selectfont{}
\caption{\textbf{A montage of the three 5-GHz EVN observations used in this work, sorted by orbital phase.} The red ellipse and red circle denote the scaled-up donor star orbit and position, as described in Figure~\ref{FigVLBAMontage}. Note that the image axis limits and the star's orbit have been scaled up by a factor of 1.7 relative to Extended Data Figure~\ref{FigVLBAMontage}, to visualise the lower-resolution $5$-GHz EVN observations at the same aspect ratio as the $8.4$-GHz VLBA observations. The contour levels denote the same level of significance as Figure \ref{FigVLBAMontage}. The grey ellipse in the bottom right of each panel shows the size of the restoring beam.  The images have been rotated by $25^{\circ}$ counter clockwise.}
\label{FigEVNMontage}
\end{figure}

\begin{figure}[h!]
\centering
\includegraphics[width=1.0\textwidth]{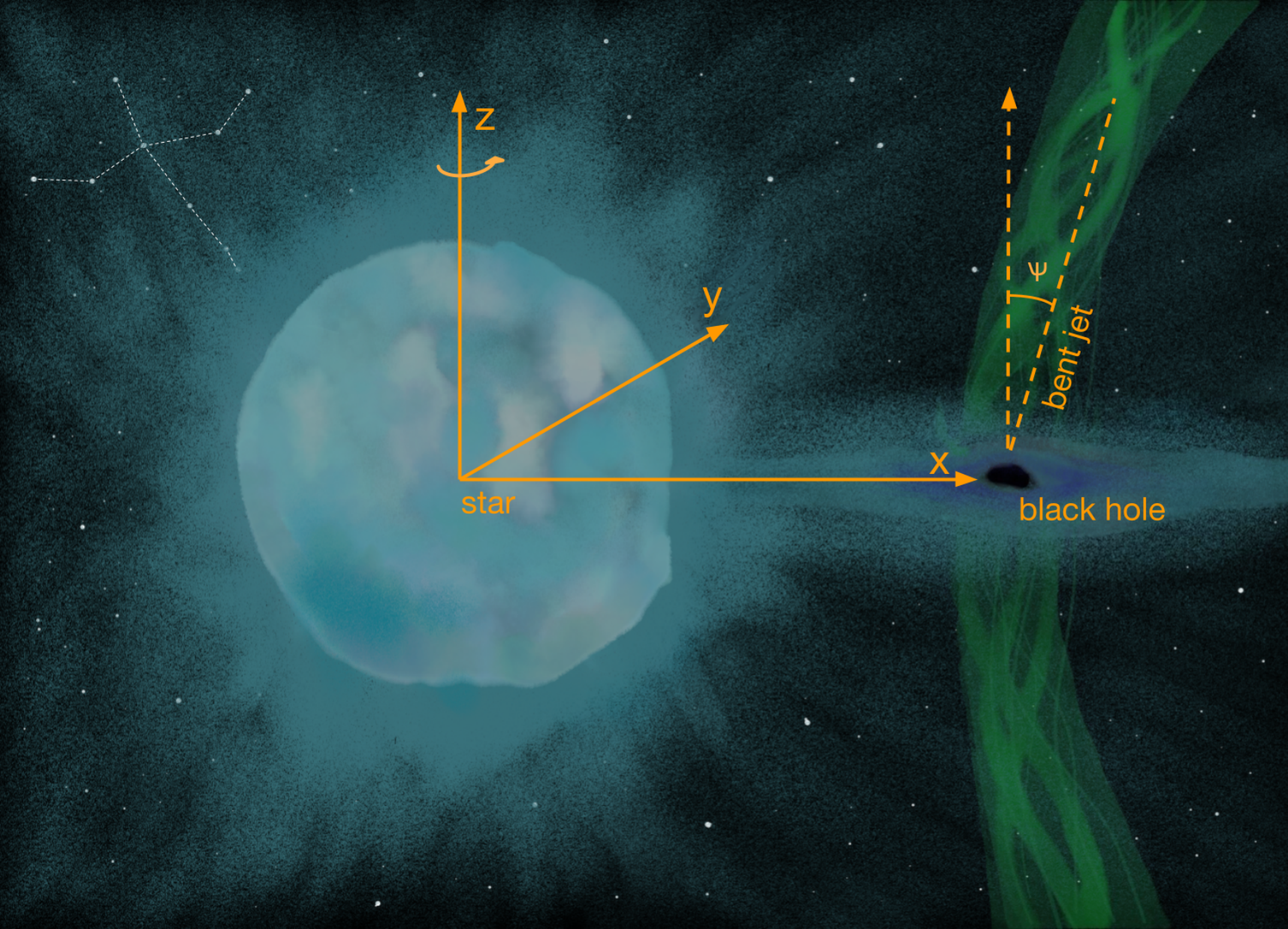}
\renewcommand{\figurename}{\textbf{Extended Data Figure}}
\linespread{1.0}\selectfont{}
\caption{\textbf{Diagram showing the non-inertial co-rotating frame used in the physical model}. In the above figure, $\psi$ is the bending angle of the jet. The misalignment angle $\theta$ is measured clockwise in the $z$-$x$ plane (in the same plane as the bending angle), with a positive angle denoting the jet pointed away from the star. For $\theta=0$ the jet would be launched along the $z$-axis.}
\label{Fig:diagram}
\end{figure}

\begin{figure}[h!]
\centering
\includegraphics[width=1.0\textwidth]{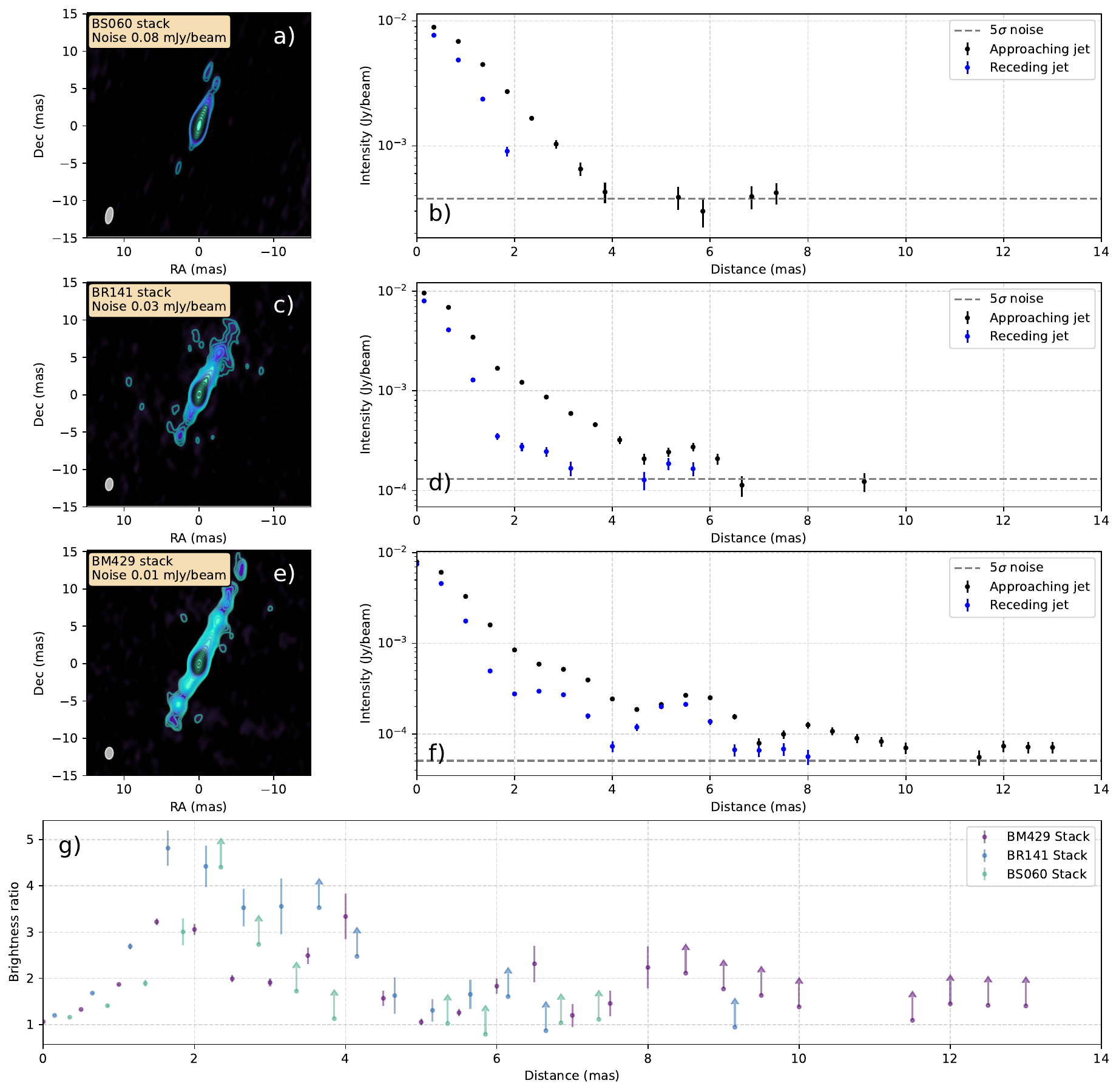}
\renewcommand{\figurename}{\textbf{Extended Data Figure}}
\linespread{1.0}\selectfont{}
\caption{\textbf{Stacked VLBA images of Cygnus X-1 from different epochs}. (a) shows a stack of three observations from 1998 (BS060[A--C]X). (c) shows a stack of four observations from 2009 (BR141[A--D]). (e) shows a stack of six observations from 2016 (BM429[B--G]). The corresponding intensity profiles along the approaching and receding jets for the stacked images are provided in (b), (d), and (f), respectively. (g) shows the corresponding measurements/limits on the brightness ratio at equal angular separation from the core for each stacked image. We find the brightness ratio to vary both by epoch and with distance from the core, ranging from $1.5-5.5$.}
\label{Fig:brightnessRatio}
\end{figure}

\begin{figure}[h!]
\centering
\includegraphics[width=\textwidth]{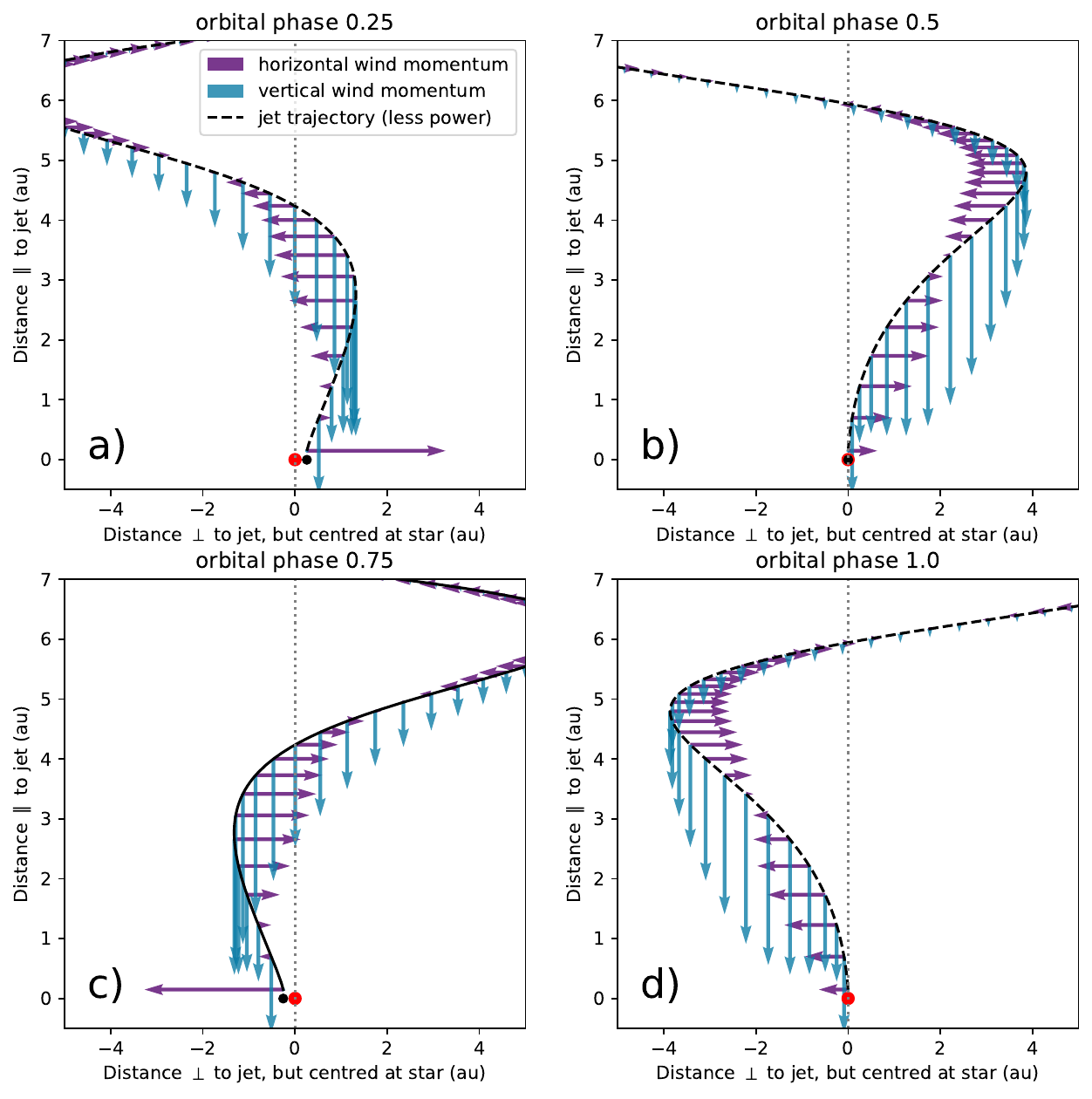}
\renewcommand{\figurename}{\textbf{Extended Data Figure}}
\linespread{1.0}\selectfont{}
\caption{\textbf{The jet trajectory calculated using momentum transfer between the jet and the wind}. We show here the trajectories for orbital phases (a) 0.00, (b) 0.25, (c) 0.50, and (d) 0.75, and provide an animation of the full orbit as a supplementary video. The reference frame is inertial and is centered on the donor star indicated by the red dot. The location of the black hole is indicated by the black circle. For calculating the jet trajectory, we assume a jet power of $5\times 10^{35}$\,ergs\,s$^{-1}$, a jet launch speed of $\beta=0.9$, a conical jet with half opening angle of $1^{\circ}$, and no misalignment between the initial jet launch direction and the binary orbit. Note that we purposefully assume a less powerful jet for better visualisation of the bent helical jet trajectory. The momentum imparted to the jet by the wind is shown parallel and perpendicular to the orbital angular momentum vector (and also the jet axis at launch) using cyan and purple arrows, respectively.}
\label{FigJetWindMomentum}
\end{figure}

\begin{figure}[h!]
\centering
\includegraphics[width=0.8\textwidth]{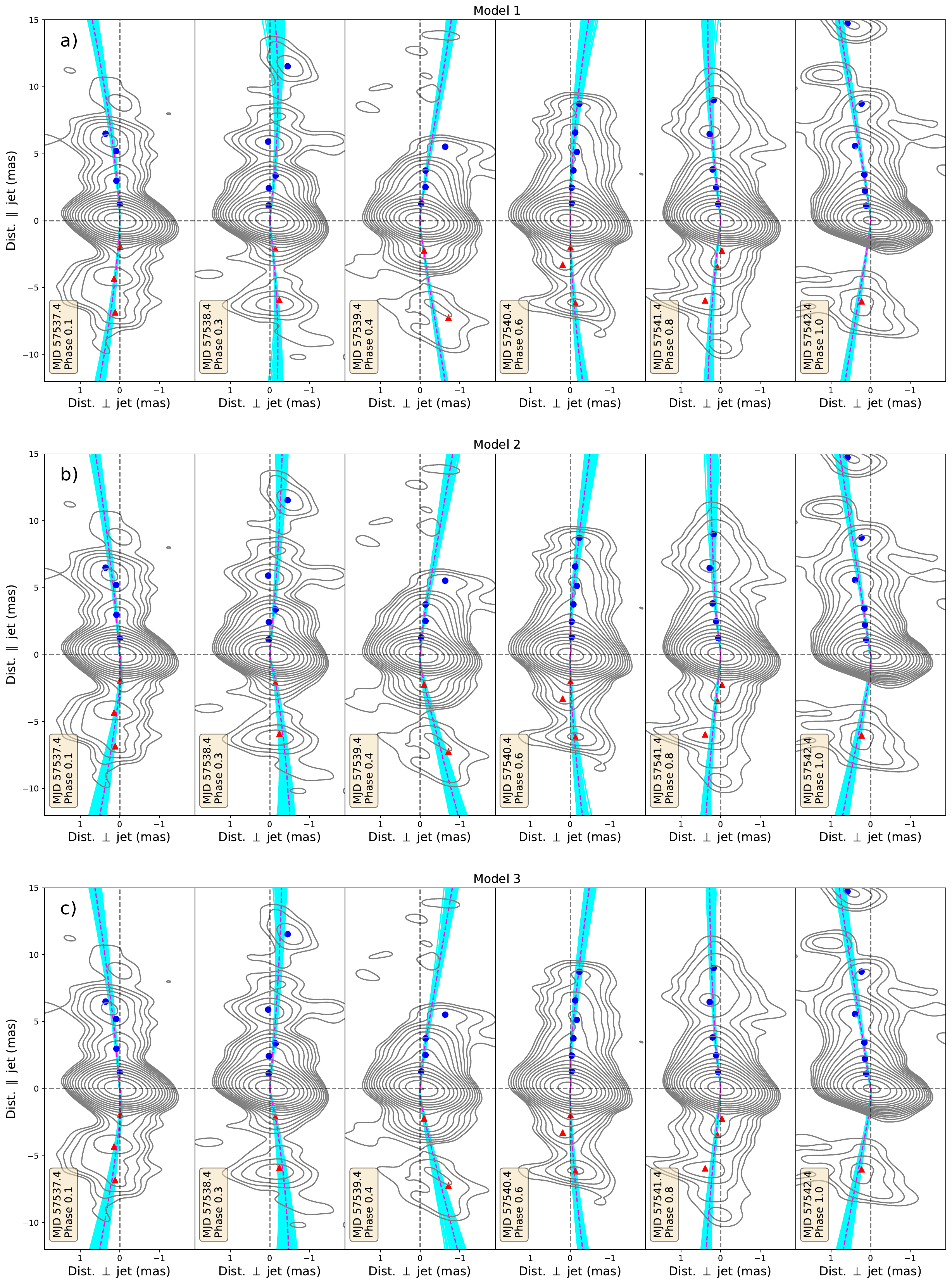}
\renewcommand{\figurename}{Extended Data Figure}
\linespread{1.0}\selectfont{}
\caption{\textbf{Fitted model trajectories for each of the VLBA observations in 2016, using the three different physical models considered in this paper.} Panel (a) shows Model 1, with a conical jet and no misalignment. Panel (b) shows Model 2, with a conical jet and misalignment. Panel (c) shows Model 3, with misalignment and a non-conical jet. As in Figure~\ref{Fig:TrajectoriesCase1}, the image from each epoch is shown with a contour plot, and the blue (red) markers show the location of the fitted point source components along the approaching (receding) jet. The dashed line shows the median fit trajectory and the cyan lines represent 200 random draws from the posterior distribution of the fit. All images have been rotated by $25^{\circ}$ counterclockwise. }
\label{Fig:allFitsTrace}
\end{figure}

\begin{figure}[h!]
\centering
\includegraphics[width=1.0\textwidth]{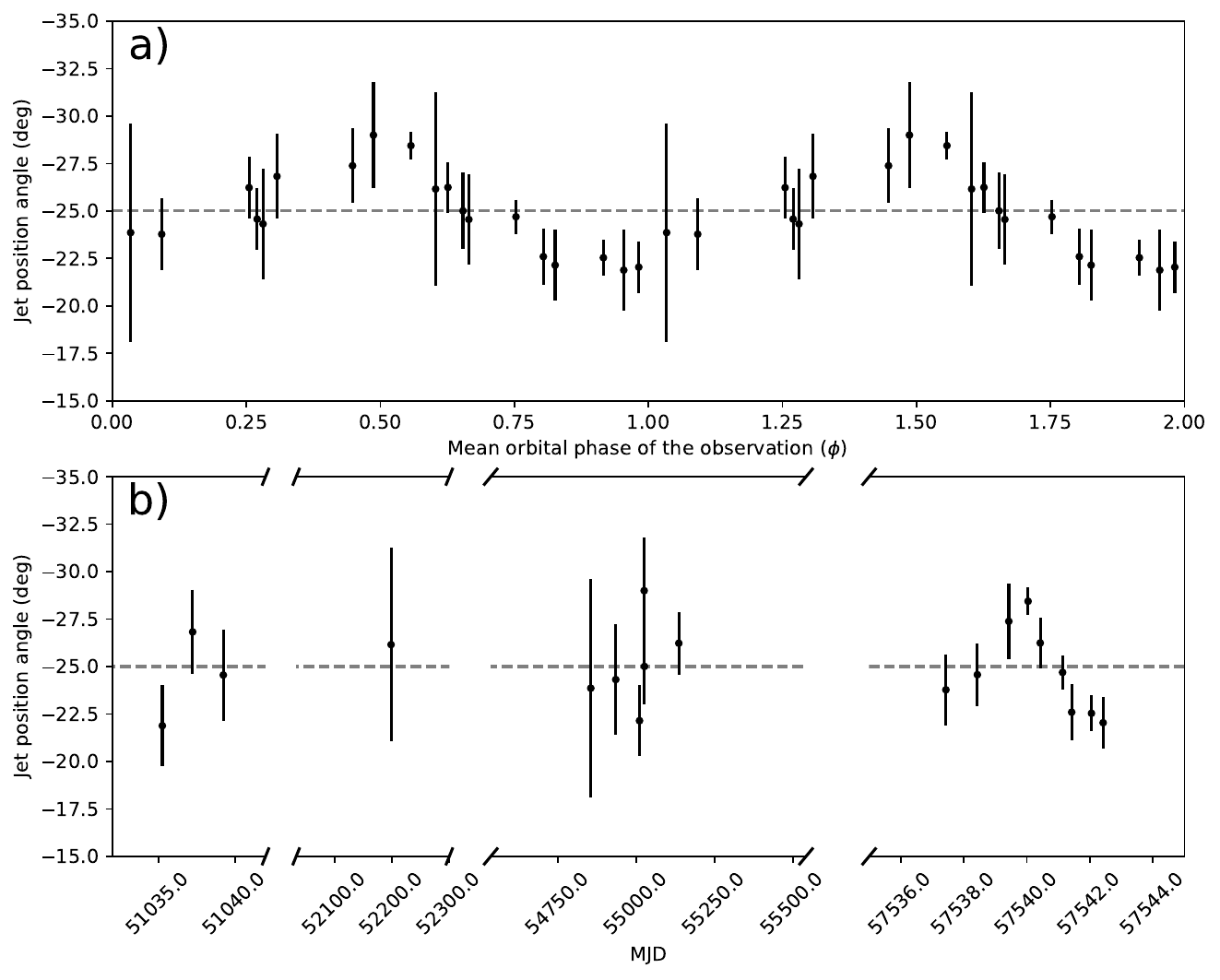}
\renewcommand{\figurename}{Extended Data Figure}
\linespread{1.0}\selectfont{}
\caption{\textbf{Position angles of jet components along the approaching jet as a function of both orbital phase and time, for all our VLBI observations from 1998--2016.} Panel (a) shows all the archival observations as a function of orbital phase, and Panel (b) shows the data as a function of time. }
\label{FigJetPA}
\end{figure}

\begin{figure}[h!]
\centering
\includegraphics[width=1.0\textwidth]{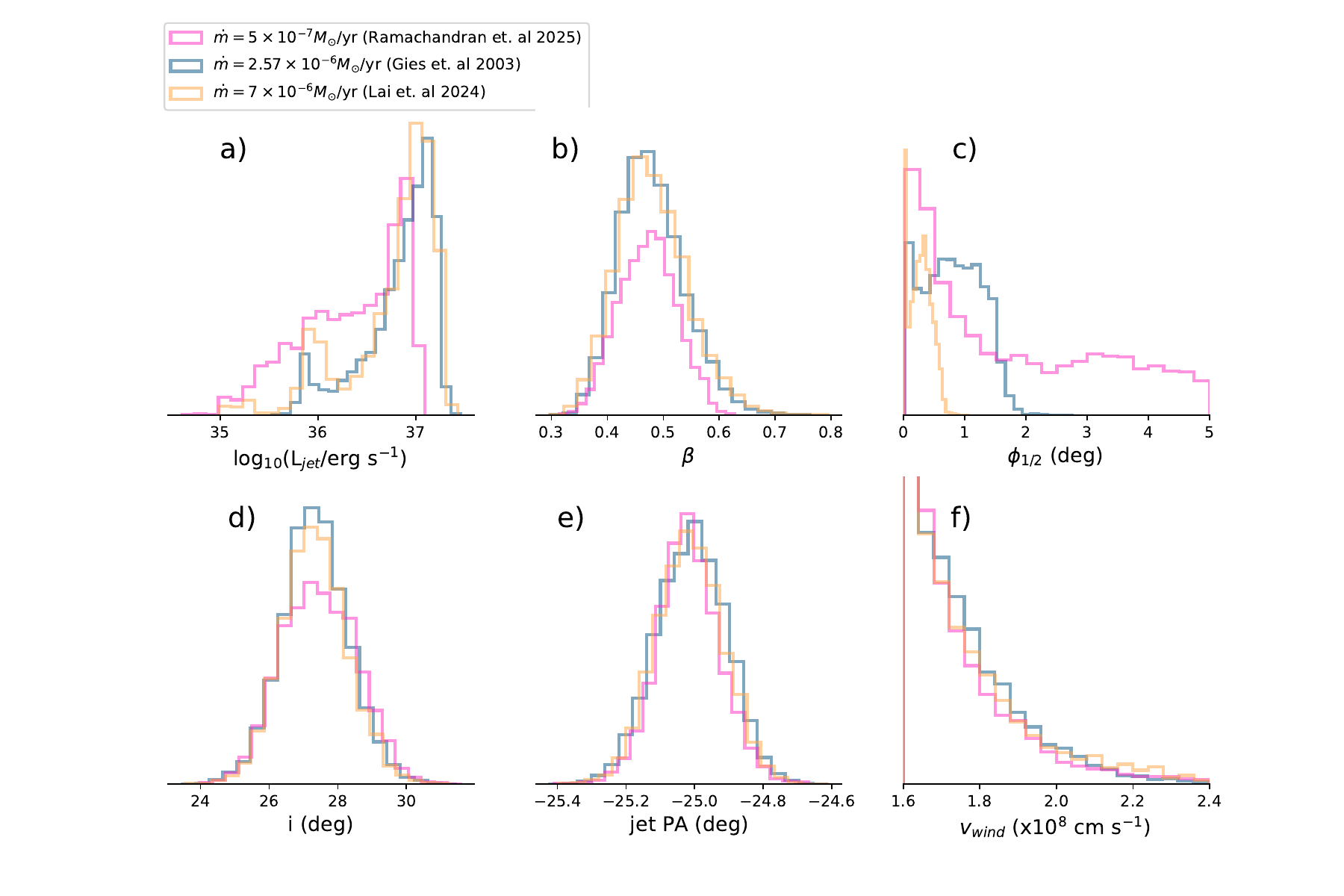}
\renewcommand{\figurename}{Extended Data Figure}
\linespread{1.0}\selectfont{}
\caption{\textbf{Histogram of model-fit traces for the different mass-loss rates.} Panels (a)-(f) show the histogram of the different fit parameters in Model 1 for different mass-loss rates. }
\label{masslossrate}
\end{figure}


\end{methods}


\bibliographystylemethods{naturemag}
\bibliographymethods{sample}

\begin{table}
\caption{Extended Data Table: VLBI observations of Cygnus X-1}\label{tab:obs}%
\small
\scalebox{0.7}{ 
\begin{tabular}{@{}ccc | ccc@{}}
    Epoch         & UTC Start        & Instrument & Epoch         & UTC Start        & Instrument \\
        Phase         & UTC End          & Freq (GHz) & Phase         & UTC End          & Freq (GHz) \\
        \hline 
        BM429B        & 2016-05-29T04:28 & VLBA       & BS060AX       & 1998-08-10T03:30 & VLBA+VLA   \\
        0.09$\pm$0.04 & 2016-05-29T16:26 & 8.4        & 0.95$\pm$0.02 & 1998-08-10T08:59 & 8.4        \\
        \hline
        BM429C        & 2016-05-30T04:25 & VLBA       & BS060BX       & 1998-08-12T03:30 & VLBA+VLA   \\
        0.27$\pm$0.04 & 2016-05-30T16:22 & 8.4        & 0.31$\pm$0.02 & 1998-08-12T07:59 & 8.4        \\
        \hline
        BM429D        & 2016-05-31T04:21 & VLBA       & BS060CX       & 1998-08-14T03:00 & VLBA+VLA   \\
        0.45$\pm$0.04 & 2016-05-31T16:19 & 8.4        & 0.66$\pm$0.02 & 1998-08-14T08:36 & 8.4        \\
        \hline
        BM429E        & 2016-06-01T04:17 & VLBA       & BS110X        & 2001-10-15T19:30 & VLBA       \\
        0.62$\pm$0.04 & 2016-06-01T16:15 & 8.4        & 0.60$\pm$0.04 & 2001-10-16T07:29 & 8.4        \\
        \hline
        BM429F        & 2016-06-02T04:13 & VLBA       & BR141A        & 2009-01-23T14:39 & VLBA       \\
        0.80$\pm$0.04 & 2016-06-02T16:11 & 8.4        & 0.03$\pm$0.03 & 2009-01-23T23:24 & 8.4        \\
        \hline
        BM429G        & 2016-06-03T04:09 & VLBA       & BR141B        & 2009-04-13T09:24 & VLBA       \\
        0.98$\pm$0.04 & 2016-06-03T16:08 & 8.4        & 0.28$\pm$0.03 & 2009-04-13T18:09 & 8.4        \\
        \hline
        RT013A        & 2016-05-31T17:30 & EVN        & BR148A        & 2009-06-28T06:00 & VLBA       \\
        0.56$\pm$0.06 & 2016-06-01T08:30 & 5.0        & 0.83$\pm$0.03 & 2009-06-28T14:29 & 8.4        \\
        \hline
        RT013B        & 2016-06-01T20:27 & EVN        & BM295A        & 2009-07-14T03:14 & VLBA       \\
        0.75$\pm$0.05 & 2016-06-02T10:06 & 5.0        & 0.65$\pm$0.02 & 2009-07-14T09:13 & 8.4        \\
        \hline
        RT013C        & 2016-06-02T18:02 & EVN        & BR141C        & 2009-07-13T03:27 & VLBA       \\
        0.92$\pm$0.05 & 2016-06-03T08:30 & 5.0        & 0.49$\pm$0.03 & 2009-07-13T12:11 & 8.4        \\
        \hline
                      &                  &            & BR141D        & 2009-10-31T20:10 & VLBA       \\
                      &                  &            & 0.25$\pm$0.03 & 2009-11-01T04:55 & 8.4        \\     
        \hline
                      &                  &            & BR141E        & 2010-01-25T13:53 & VLBA       \\
                      &                  &            & 0.57$\pm$0.04 & 2010-01-25T23:56 & 8.4        \\             
\end{tabular}
} \setcounter{table}{1} 
\linespread{1.0}\selectfont{}\caption*{\textbf{Extended Data Table 1: Log of all VLBI observations used.} The left column lists the 9 VLBI observations that closely sampled the single $5.6$ day binary orbit in 2016. The remaining archival observations that we used are listed in the right column. Note that the epoch BR141E did not show a resolved jet and hence is not used in this study. Orbital phases are calculated using the spectroscopic ephemeris of \cite{brocksopp1999}.
}
\end{table}

\begin{table}
\caption{Extended Data Table: Parameters from physical model}\label{tab:modelfits}%
\scalebox{0.7}{ 
\begin{tabular}{@{}ccccc@{}}
\scriptsize 
    Parameter & & model 1 & model 2 & model 3 \\
    \hline \hline
    $\log_{10}$(L$_{\rm jet}$/erg\,s$^{-1}$) & prior & $\mathcal{U}(33-38)$ & $\mathcal{U}(33-38)$ & $\mathcal{U}(33-38)$\\
     & posterior & $36.94_{-0.49}^{+0.22}$ & $37.28_{-0.23}^{+0.13}$ & $37.26_{-0.25}^{+0.13}$\\
    \hline
    $\beta$ & prior & $\mathcal{U}(0.01 - 0.99)$ & $\mathcal{U}(0.01 - 0.99)$
  & $\mathcal{U}(33-38)$\\
   & posterior & $0.47_{-0.05}^{+0.06}$  & $0.68_{-0.13}^{+0.08}$ & $0.66_{-0.09}^{+0.08}$\\
    \hline
    $i$ & prior & $\mathcal{N}(27.5 \pm 1)$ & $\mathcal{N}(27.5 \pm 1)$ & $\mathcal{N}(27.5 \pm 1)$\\
    (degrees)& posterior & $27.32_{-0.91}^{+0.96}$ & $27.49_{-0.99}^{+0.98}$ & $27.82_{-0.75}^{+0.83}$\\
    \hline
    $\phi_{1/2}$ & prior & $\mathcal{U}(0.001 - 5)$ & $\mathcal{U}(0.001 - 5)$ & $\mathcal{U}(0.001 - 5)$\\
    & posterior & $0.8_{-0.5}^{+0.5}$ & $1.8_{-0.8}^{+0.7}$ & $1.6_{-0.7}^{+0.6}$\\
    \hline
    $\theta$ & prior & & $\mathcal{U}(0 - 30)$ & $\mathcal{U}(0 - 30)$\\
    (degrees)& posterior & & $5.2_{-1.1}^{+1.0}$ & $4.9_{-0.93}^{+0.81}$ \\
    \hline
    az & prior & & $\mathcal{U}(0 - 1)$ & $\mathcal{U}(0 - 1)$\\
    & posterior & & $0.34_{-0.02}^{+0.03}$ & $0.34_{-0.02}^{+0.03}$\\
    \hline
    Jet PA & prior & $\mathcal{N}(0\pm5)$ & $\mathcal{N}(0\pm5)$ & $\mathcal{N}(0\pm5)$\\
    (degrees)& posterior & $-25.0_{-0.1}^{+0.1}$ & $-27.8_{-0.4}^{+0.5}$ & $-27.7_{-0.4}^{+0.5}$ \\
    \hline
    eps & prior & & & $\mathcal{U}(0.1 - 2)$\\
    & posterior & & & $1.02_{-0.08}^{+0.08}$ \\
    \hline
    $v_{\rm wind}$ & prior & $\mathcal{U}(1.6-2.4)$&$\mathcal{U}(1.6-2.4)$ & $\mathcal{U}(1.6-2.4)$\\
    ($\times10^{6}$\,m\,s$^{-1}$) & posterior & $\le1.87$ & $\le 1.98$ & $\le 1.98$\\
    \hline \hline
    
\end{tabular}
} \setcounter{table}{2} 
\linespread{1.0}\selectfont{}\caption*{\textbf{Extended Data Table 2: Priors and posteriors of the physical models}. Model 1 represent a conical jet initially aligned with the binary orbit. Model 2 represents a misaligned conical jet, and Model 3 represents a misaligned jet that can deviate from a conical geometry. In the above table, $\mathcal{U}$ represents a uniform prior with the range of the bounds mentioned within parentheses. $\mathcal{N}$ represents a normal prior, with the corresponding mean and standard deviation given in parentheses. For the posteriors we mention the $68\%$ credible intervals.  }
\end{table}

\begin{table}
\caption{Extended Data Table: Spectral index from VLA.}\label{tab:vla}%
\scalebox{0.7}{ 
\begin{tabular}{ccc}
\scriptsize 
    Date & Orbital Phase & Spectral Index \\
        \hline
        
        2016 May 30 & 0.284$\pm$0.003 & -$0.082 \pm 0.018$ \\
        2016 May 31 & 0.418$\pm$0.015 & $0.009 \pm 0.022$ \\
        2016 June 1 & 0.601$\pm$0.003 & $0.055 \pm0.016$ \\
        2016 June 2 & 0.766$\pm$0.003 & $0.040 \pm 0.016$ \\
        2016 June 3 & 0.957$\pm$0.015 & -$0.141 \pm 0.028$ \\
        2016 June 8 & 0.842$\pm$0.003 & $0.026 \pm 0.011$ \\
        \hline
        
\end{tabular}
} \setcounter{table}{3} 
\linespread{1.0}\selectfont{}\caption*{\textbf{Extended Data Table 3: The spectral index of Cygnus X-1 measured from our 2016 VLA observations}. The spectral index was determined by fitting a power law to the measured flux density of the unresolved source in each of the spectral windows in the 5.25 and 7.45 GHz VLA basebands. }
\end{table}

\begin{table}
\caption{Extended Data Table: Cygnus X-1 VLBI flux densities}\label{tab:fluxdensities}%
\scalebox{0.7}{ 
\begin{tabular}{lcccc}
\scriptsize 
        Epoch & Core flux density & Total flux density & Noise & Frequency \\
        & (mJy) & (mJy) & (mJy\,beam$^{-1}$) & (GHz) \\
        \hline
        BM429B & 5.44$\pm$0.55 & 8.11$\pm$0.57 & 0.02 & 8.4 \\
        BM429C & 6.41$\pm$0.64 & 9.36$\pm$0.67 & 0.02 & 8.4\\
        BM429D & 6.85$\pm$0.69 & 9.61$\pm$0.70 & 0.02 & 8.4\\
        BM429E & 8.03$\pm$0.80 & 11.19$\pm$0.82 & 0.02 & 8.4\\
        BM429F & 7.92$\pm$0.79 & 11.12$\pm$0.81 & 0.02 & 8.4\\
        BM429G & 6.52$\pm$0.66 & 9.81$\pm$0.68 & 0.02 & 8.4\\
        RT013A & 6.39$\pm$0.64 & 8.10$\pm$0.65 & 0.02 & 5.0\\
        RT013B & 7.24$\pm$0.72 & 8.80$\pm$0.73 & 0.02 & 5.0\\
        RT013C & 8.05$\pm$0.81 & 9.67$\pm$0.81 & 0.02 & 5.0\\
        BS060AX & 8.17$\pm$0.84 & 12.20$\pm$0.92 & 0.07 & 8.4\\
        BS060BX & 11.16$\pm$1.16 & 15.24$\pm$1.23 & 0.12 & 8.4\\
        BS060CX & 11.48$\pm$1.16 & 14.62$\pm$1.19 & 0.09 & 8.4\\
        BS110X & 16.84$\pm$1.71 & 17.81$\pm$1.73 & 0.16 & 8.4\\
        BR141A & 7.92$\pm$0.80 & 10.91$\pm$0.83 & 0.04 & 8.4\\
        BR141B & 8.42$\pm$0.85 & 11.01$\pm$0.88 & 0.04 & 8.4\\
        BR148A & 12.89$\pm$1.29 & 16.05$\pm$1.32 & 0.03 & 8.4\\
        BM295A & 12.97$\pm$1.32 & 19.67$\pm$1.42 & 0.10 & 8.4\\
        BR141C & 11.52$\pm$1.17 & 18.47$\pm$1.24 & 0.05 & 8.4\\
        BR141D & 7.93$\pm$0.81 & 10.04$\pm$0.83 & 0.04 & 8.4\\
        BR141E & 9.95$\pm$1.00 & 9.95$\pm$1.00 & 0.05 & 8.4\\
        \hline
                
\end{tabular}
} \setcounter{table}{3} 
\linespread{1.0}\selectfont{}\caption*{\textbf{Extended Data Table 4: Measured flux densities of Cygnus X-1 during the VLBA observations analysed in this work.} The core flux density was determined by fitting a two-dimensional Gaussian to the brightest point in the image, whereas the total flux density also includes the downstream emission from the jets. }
\end{table}

\noindent{\bf Extended Data}

\clearpage
\noindent
{\bf Supplementary Video: }

\end{document}